\documentclass[aps,
onecolumn,
superscriptaddress
]{revtex4-2}

\usepackage{amsmath,amssymb,physics,bm}
\usepackage{graphicx, microtype,siunitx} 
\usepackage[colorlinks=true,linkcolor=blue,citecolor=blue,urlcolor=blue]{hyperref}
\usepackage[all]{hypcap} 
\usepackage[colorinlistoftodos]{todonotes}

\usepackage{mathptmx}
\newcommand{\nvec}[0]{{\bf n}}
\newcommand{\rvec}[0]{{\bf r}}
\newcommand{\Evec}[0]{{\bf E}}

\newcommand{\bs}[0]{\boldsymbol\sigma}
\usepackage{dsfont}

\begin{document}

\title[]{Liquid crystal skyrmions as elastic multipoles}

\author{Allison W. Teixeira}
\email{allwtx@gmail.com}
\affiliation{Centro de Física Teórica e Computacional, Faculdade de Ciências, Universidade de Lisboa, 1749-016 Lisboa, Portugal.}%Lines break automatically or can be forced with \\
 \affiliation{Departamento de Física, Faculdade de Ciências, Universidade de Lisboa, P-1749-016 Lisboa, Portugal.}
 \affiliation{Instituto de Alta Investigación, Universidad de Tarapacá, Casilla 7D, Arica, Chile}

\author{Cristóvão S. Dias}
\email{csdias@fc.ul.pt}
\affiliation{Centro de Física Teórica e Computacional, Faculdade de Ciências, Universidade de Lisboa, 1749-016 Lisboa, Portugal.}%Lines break automatically or can be forced with \\
 \affiliation{Departamento de Física, Faculdade de Ciências, Universidade de Lisboa, P-1749-016 Lisboa, Portugal.}

\author{Mykola Tasinkevych}
\email{mykola.tasinkevych@ntu.ac.uk}
\affiliation{SOFT Group, School of Science and Technology, Nottingham Trent University, Clifton Lane, Nottingham NG11~8NS, United Kingdom.}
\affiliation{International Institute for Sustainability with Knotted Chiral Meta Matter, Hiroshima University, Higashihiroshima 739-8511, Japan.}

\date{\today}

\begin{abstract}

\noindent{\bf \Large Abstract}

\noindent Solitons in liquid crystals are spatially localised stable configuration of the liquid crystal orientational order parameter that exhibit emergent particle-like properties such as mutual interaction,  translational motion and reconfigurable self-asembling behaviour in oscillating electric fields. Understanding solitons' out-of-equilibrium interactions remains a significant challenge. In this study, we present a minimalist model capable to address this challenge. We derive analytically elastic multipole interaction potentials between two solitons assuming two dimensional LC configurations. We determine the soliton's elastic dipole and quadrupole moments based on the equilibrium director textures obtained as minimizers of the Frank-Oseen free energy functional.
Incorporating the time dependence of the soliton's multipole moments in response to electric field oscillations, we carry out particle-based simulations that demonstrate the formation of soliton chains and two dimensional clusters by varying the amplitude and/or the frequency of the electric field, in agreement with experiments. The presented formalism may serve as a benchmark model for future theoretical developments incorporating, e.g. higher order elastic multipoles or/and many body interactions.
  
\end{abstract}

\maketitle

\noindent{\Large \bf Introduction}\newline

\noindent Liquid crystals (LCs) are fascinating soft materials which enable novel reconfigurable photonic and nonlinear optics applications \cite{Liu2014,mundor:2015,Zhang2015,hess:2020,papic:2021,Meng2023} as well as composite nanomaterials with on demand  properties tuneable by electric field or temperature \cite{li:2017,Mertelj2013,Mundoor:2016,Mundoor2021}. On the other hand, basic investigations of the behaviour of LCs under different conditions provide insights into other branches of physics such cosmological theories of the early universe or superfluids.
For example, nematic LCs have been used to demonstrate the validity of the prediction of the Kibble-Zurek mechanism regarding the scaling behaviour of the defect density during continuous phase transitions \cite{fowler:2017}. It also was shown that LC topological defect lines can continuously bend beams of light, offering a testbed for probing the interaction of light with, e.g. cosmic strings \cite{Meng2023}. LCs also offer a unique opportunity to explore various manifestations of topology in a context of condensed matter physics. Indeed, chiral LCs enable a large variety of thermodynamically stable solitons which are topologically protected localized distortions of the director field. For instance, two-dimensional (2D) skyrmions \cite{Fukuda2011,Ackerman2014,Posnjak2016}, and three-dimensional (3D) hopfions \cite{Ackerman2017b,voinescu:2020}, torons \cite{Smalyukh2010,guo:2016}, heliknotons, \cite{tai:2019} and möbiusons \cite{Zhao2023} have been reported experimentally. 

LC solitons exhibit emergent particle-like properties such as a translational motion in time-dependent electric fields \cite{Ackerman2017,Sohn2019} and effective interactions that can be easily tuned in strength or switched from attractive to repulsive \cite{Sohn2019}. Reference~\cite{Foster2019} used a  superposition of two axially symmetric {\it Ansätze} for the director field of 2D skyrmions and found isotropic repulsion. The soliton interactions are intrinsically out of equilibrium when the period of driving electric field is shorter than the relaxation time of the LC director, resulting in very rich collective dynamics with reconfigurable transient soliton aggregates \cite{sohn2020}.  Despite  extensive body of experimental research, understanding the many-body dynamics of LC solitons remains an outstanding challenge. Existing numerical investigations are limited to small number of skyrmions and exploit pure relaxational dynamics to track the temporal evolution of the director field \cite{Ackerman2017, Sohn2019}. 

In our earlier work \cite{teixeira2024}, we have developed a particle-based model for motion of a skyrmion driven by electric field. The model introduces coarse-grained external time-dependent forces describing the tendency of skyrmions to move forward and backward when the electric field is turned on and off. The functional form of the forces is motivated by results of a fine-grained description based on numerical minimization of the Frank-Oseen free energy. The coarse-grained model successfully reproduces experimental results \cite{Ackerman2017} on the  reversal of the skyrmion velocity depending on the frequency of the electric field. 

Here, we generalise this approach by introducing soliton interactions to the particle-based model of \cite{teixeira2024}. The interactions are considered at a pairwise level and coarse-grained analytical expression are derived using 2D elastic multipole expansion developed by Pergamenshchik and Uzunova  \cite{Pergamenshchik2007-gi} for nematic colloids. We calculate the elastic dipole and quadrupole moments at static conditions from equilibrium director fields around skyrmions with and without electric field applied. At time-varying electric fields, we assume an exponential relaxation dynamics of multipole moments resulted form the continuous morphing of the LC director when the electric field is turned on and off. Molecular dynamics simulations show formation of soliton chains and 2D clusters driven by the interplay between dipole-dipole and quadrupole-quadrupole interactions, which vary with changing electric field and the cholesteric pitch. We estimate a ground state chain-cluster configuration diagram highlighting the transitions between chains and clusters by varying the amplitude and frequency of the field, in a qualitative agreement with the experimental observations \cite{Sohn2019}. The presented model may serve as a benchmark for future developments of more accurate descriptions of solitons collective behaviour taking into account, e.g. many body effects or/and multipoles of higher order.  \newline  

\noindent {\Large \bf Model}\newline

\noindent {\bf Frank-Oseen free energy}\newline

\begin{figure}[ht]
  \centering
  \includegraphics[width=0.6\columnwidth]{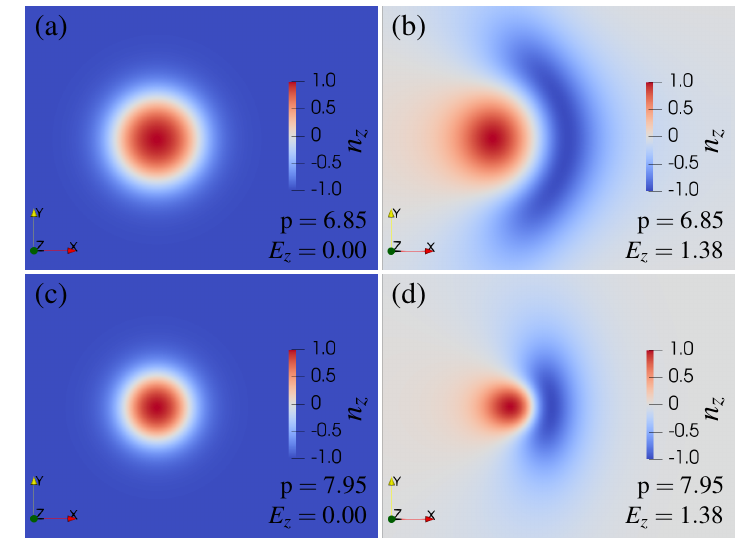} 
  \caption{{\bf Equilibrium director configurations for skyrmions and bimerons.} Color coded $z-$component $n_{z}$ of the equilibrium director field obtained by numerical minimization of 2D Frank-Oseen free energy functional. (a) Axially symmetric skyrmion configuration obtained at zero external electric field and $\mathrm{p}=6.85$, and (b) asymmetric bimeron configuration at $\mathrm{p}=6.85$ and $E_z  = 1.38$, recall that the electric field used here has the form ${\bf E} = E_z \hat {\bf z}$. (c) demonstrates the effect of increasing the pitch on the spatial extent of the skyrmion; $\mathrm{p} = 7.95$, $E_z  =0$. Increasing the pitch also decreases the spatial region occupied by the asymmetric bimeron texture (d), which is obtained at $\mathrm{p} = 7.95$ and $E_z = 1.38$. The bimeron is unstable for $\mathrm{p} \ge 8.04$.  }
  \label{SkBim}
\end{figure}

\noindent We consider here a two-dimensional system where the director $\nvec = (n_x,n_y,n_z)^{\mathrm T}$ field varies in the $(x,y)$ plane only; $(...)^{\mathrm T}$ denotes transpose operation.  At zero electric field this corresponds to a three-dimensional system which is translationally invariant along the skyrmion's symmetry axis, see Figs.~\ref{SkBim}(a) and (c). This analogy, however, does not apply to electric field-deformed configurations, as shown in Figs.~\ref{SkBim}(b) and (d). The Frank-Oseen elastic free energy per unit length reads:
\begin{equation}
       \label{eq:free_energy}
    F=\int_{\mathds{R}^{2}} \biggl(  \frac{k_{1}}{2}
    \left( {\bf \nabla} \cdot {\bf n} \right)^{2} 
    + \frac{k_{2}}{2} \left[ {\bf n} \cdot \left( {\bf \nabla \times {\bf n}} \right) + \frac{2 \pi}{\mathrm{p}} \right]^{2}
     + \frac{k_{3}}{2} \left| {\bf n} \times \left(
    {\bf \nabla} \times {\bf n} \right) \right|^{2} - 
    \frac{\varepsilon_0 \Delta\varepsilon}{2} ({\bf E} \cdot {\bf n})^2 \biggl) d^2r.
\end{equation}
where ${\bf n}$ is the nematic director field, $k_1, k_{2}$ and $k_3$ are positive elastic constants describing splay, twist and bend director distortions, respectively, $\mathrm{p}$ is the cholesteric pitch. External electric field ${\bf E}$ couples to the nematic director according to the last term under the integral, where $\varepsilon_0$ is the vacuum permittivity and $\Delta\varepsilon$ is the dielectric anisotropy. The last term in the integrand of Eq.~(\ref{eq:free_energy}) approximates the local electric field by the constant external field. This approximation is valid for weak fields which is the case of the experiments reported in \cite{Ackerman2017,sohn2020}. In this study we assume ${\bf E} = E_z \hat {\bf z}$, where $\hat {\bf z}$ is the unit vector in the $z-$direction.
 
 In experimental realisations, liquid crystal is place between parallel plates which impose perpendicular boundary conditions on the LC director. In our 2D model, this effect is accounted for by an additional free energy density  term
\begin{align}
    u_{w} = -\frac{W_{0}}{2} \left( n_z \right)^{2},
    \label{fw}
\end{align}
where $W_{0}$ is the effective anchoring strength which favours director alignment in the $z-$direction, mimicking homeotropic anchoring conditions at the cell surfaces of a three-dimensional system. \textcolor{black}{This is a standard approach to account for a boundary-induced anisotropy through an effective bulk electric field in 2D system, see \cite{Duzgun2018}, for a detailed discussion.}

The equilibrium director configurations are obtained numerically by minimising the free energy \eqref{eq:free_energy} augmented with the effective anchoring free energy density in Eq.~\eqref{fw} (see Methods for details). Figure~\ref{SkBim} shows typical equilibrium director configurations corresponding to different values of the pitch and the electric field. Rotationally symmetric skyrmion configuration emerges when no electric field is applied (Figs.~\ref{SkBim}(a) and \ref{SkBim}(c)). When electric field is applied, the skyrmion morphs into an asymmetric configuration (Figs.~\ref{SkBim}(b) and \ref{SkBim}(d)). These structures were first observed in LCs \cite{Ackerman2015} and were called  "toron-umbilical pair", which is equivalent to  "bimeron" configuration known in magnetic skyrmion literature \cite{Li2020}.

The region containing a solitonic texture, either skyrmion or bimeron, is characterised by strong variations of the director field.
The size of this region depends on the pitch and the strength of the electric field, as shown in Fig.~\ref{SkBim}. When the pitch increases, the size of the solitonic textures decreases and eventually disappears at a specific critical pitch, in agreement with previous results \cite{sohn2019a}. Far from the solitons, the director field is approximately uniform and approaches the far field director ${\bf n}_{0}$ as distance from the soliton grows. The long distance perturbation of ${\bf n}_{0}$ due to the presence of a soliton, which is explored in detail in the following sections,  depends on the texture's size and symmetry. 

For skyrmions, ${\bf n}_{0}=\hat{\bf z}$ and the small director perturbation are in the $(x,y)$ plane. On the other hand, the far field director and the perturbation due to the bimerons both are within the $(x,y)$ plane and perpendicular to each other. 
To get insights into the relation between the director perturbations and emerging paiwise soliton interactions, we applied the colloidal nematostatics method proposed by Pergamenshchik and Uzunova in \cite{Pergamenshchik2007-gi} for dispertions of colloidal particles in LCs. \newline

\noindent {\bf Far-field perturbations induced by skyrmions} \newline

\noindent We present the director field around a soliton in the following form ${\bf n} = {\bf n}_{0} + \delta{\bf n}$. For the case of a symmetric skyrmion ${\bf n}_{0}=(0,0,n_{z}=1)^{\mathrm T}$ and  $\delta{\bf n} = (n_{x}({\bf r}),n_{y}({\bf r}),0)^{\mathrm T}$ where  $n_{x}({\bf r}),n_{y}({\bf r})\ll 1$. We note that the normalisation condition on ${\bf n}$  is satisfied to linear order in $n_x,n_y$, i.e. ${\bf n}\cdot{\bf n} = 1 + {\cal O}(n_x^2,n_y^2)$.   
The contribution of the electric field term in Eq.~(\ref{eq:free_energy}) is zero once the electric field is zero, and Eq.~(\ref{fw}) at leading order contributes to the free energy a constant since $n_z^2 = 1$, and therefore can be neglected. With these, and adopting one elastic constant approximation, $k_{1}=k_{2}=k_{3}=k$, the free energy at leading order can be written as 
\begin{align}
    U =
    \frac{k}{2} \int_{\mathds{R}^{2}}
    \left( {\bf \nabla} n_{\mu} \cdot {\bf \nabla} n_{\mu} \right)d^2r,
    \label{FrankOseen_2D_Sk_one_cst}
\end{align}
where a summation over $\mu={x,y}$ is implicitly assumed. The Euler-Lagrange equations corresponding  to free energy (\ref{FrankOseen_2D_Sk_one_cst}) are the Laplace's equations for each perturbation component $\Delta n_{\mu} = 0$. Assume that $n_{\mu}$ is known on a closed contour $\zeta$ which encompasses the skyrmion texture completely  and that the director perturbations outside $\zeta$ are small. In other words $\zeta$ separates the inner region $\zeta_<$ with large director distortions from the outer region $\zeta_>$ where the distortions are small and governed by the Gaussian theory \eqref{FrankOseen_2D_Sk_one_cst}. 
In this case, the outer solutions for $n_{\mu}({\bf r})$, where ${\bf r}\in \zeta_>,$ can be written by using 2D Green's function $G({\bf r}',{\bf r})$ of the Laplace's equation, where $G({\bf r}',{\bf r})=0$ for ${\bf r}' \in \zeta$, in the following form
\begin{align}
    n_{\mu}({\bf r}) = \int_{\zeta} n_{\mu}({\bf r}') \left( {{\bf \bs}} \cdot {\bf \nabla}_{\bf r'} \right) G({\bf r}',{\bf r}) d{\bf r}'.
    \label{nmu}
\end{align}
 ${{\bf \bs}}$ above denotes the inward unit vector normal to the contour $\zeta$. 

For the case of the asymmetric bimeron texture we assume, without loss of generality, that the far-field director ${\bf n}_0 = (0,n_y=1,0)^{\mathrm T}$ and $\delta{\bf n} = (n_{x},0,0)^{\mathrm T};~n_{x}\ll 1$. This is valid for strong enough electric fields applied along $\hat{\bf z}$, which due to the negative dielectric anisotropy force the director field to the $(x,y)$ plane.   
Then, similar to the case of the skyrmion, $n_x$ for the bimeron also obeys the Laplace's equation and can be written in the form of Eq.~(\ref{nmu}). \newline

\noindent {\Large \bf Results} \newline %label sec 2

\noindent {\bf The Multipole expansion}\newline

\noindent We assume that the contour $\zeta$, which encloses the topological texture of a soliton, is a circle of radius $a$, and that the distribution of the perturbations $n_\mu$ on $\zeta$ is known, e.g. by numerical solution of the full non-linear problem. It is important to emphasize that $a$ must be large enough such that $n_\mu\ll 1$ both at $\zeta$ and in $\zeta_>$. This construction allows to treat solitons as effective colloidal disks with Dirichlet's boundary conditions which are immersed in a liquid crystal host whose distortions are weak and governed by the Gaussian free energy (\ref{FrankOseen_2D_Sk_one_cst}).
We calculate the perturbations (\ref{nmu}) in $\zeta_>$ by using the Green's function obtained with the help of the method of images, (see Supplementary Note 1 for details).
The result for the perturbations can be written in a form of a multipole expansion 
\begin{align}
    n_{\mu}(r,\theta) = q^{(\mu)} + 2 \frac{{\bf p}^{(\mu)} \cdot {\bf r}}{r^{2}} + 4 \frac{1}{r^{4}} {\bf Q}^{(\mu)} : {\bf r} : {\bf r}+...,
    \label{perturbation}
\end{align}
\begin{align}
    q^{(\mu)} = \frac{1}{2 \pi} \int_{0}^{2 \pi} n_{\mu}(a,\theta') d\theta', \quad
    p_{\alpha}^{(\mu)} = \frac{a}{2 \pi} \int_{0}^{2 \pi} n_{\mu}(a,{\theta'}) \sigma_{\alpha} d \theta',
    \label{dipolemoment}
\end{align}
and 
\begin{align}
    Q_{\alpha \beta}^{(\mu)}= \frac{a^{2}}{4 \pi} \int_{0}^{2 \pi} n_{\mu}(a,{\theta'}) \left(2 \sigma_{\alpha} \sigma_{\beta} -\delta_{\alpha \beta} \right) d\theta',
    \label{quadrupole}
\end{align}
where we introduce polar coordinates $(x,y)=(r\cos\theta,r\sin\theta)$, ${\bf p}^{(\mu)} = (p_{x}^{(\mu)},p_{y}^{(\mu)})$,
and ${\bf Q}^{(\mu)} : {\bf r} : {\bf r} = Q_{\alpha \beta}^{(\mu)} r_{\alpha} r_{\beta}$; all the Greek indices run over $x, y$.
Recall that $\sigma_{\alpha}$ denotes components of the unit normal vector pointing to the center of the disk containing the solitonic texture.

For three-dimensional colloidal systems, the analogy between electrostatics and nematostatics was established where the director components on a spherical surface which surrounds a colloidal particle are interpreted as the surface elastic charge density \cite{Pergamenshchik2007-gi,PhysRevE.76.011707,PhysRevE.79.021704,doi:10.1080/15421400903060284}. 
In the present case the director components $n_{\mu}(a,\theta)$ on the contour of the disk may be interpreted in a similar way, such that $q^{(\mu)}$, $\bf{p}^{(\mu)}$ and $\bf{Q}^{(\mu)}$ are equivalent to the electric charge, dipole and quadrupole moments in two dimensions, respectively.
It is worth noticing that multipoles with different index $\mu$ are independent and do not interact with each other. The combined multipoles are often called {\it dyad} \cite{brochard1970theory,Pergamenshchik2007-gi}. \newline

\noindent {\bf Elastic torque on an asymmetric skyrmion}\newline %label skyrmion_far_field_interaction

\noindent In the next section we will shown that for the bimeron texture, the elastic dipole  in equilibrium is oriented parallel to the far field director, both of which are perpendicular to the vector ${\bf d}$ connecting the preimages of the south and north poles of the order parameter manifold, which is a two-dimensional sphere for a vectorized director field \cite{tai:2024}. Experiments on systems with multiple solitons show  that at early times, just after the electric field has been turned on, ${\bf d}_i$ of individual solitons are aligned randomly and in a course of time the vectors ${\bf d}_i$ align along a common direction perpendicular to the global far field. This observation shows that if the elastic dipole of a skyrmion is misaligned with the global far field, there must exist an effective torque which drives the elastic dipole to the equilibrium orientation. The torque stems from soliton-soliton interactions and also from the interaction of a soliton and the far field director. Here, we discuss an approximate expression which quantifies the later effect. 

To this end, we evaluate the variation of the free energy \eqref{FrankOseen_2D_Sk_one_cst} resulting from the "rigid body" rotation of the a disk enclosing the solitonic texture about the $z-$ axis. The free energy associated with the outer region of an effective "colloidal particle" $i$  can be calculated by substituting Eq.~(\ref{nmu}) into Eq.~(\ref{FrankOseen_2D_Sk_one_cst}), yielding
\begin{align}
    U_{i} & =
     \frac{k}{2} \int_{\zeta} \int_{\zeta}
     n_{\mu}({\bf r}')
     u_{i}({\bf r}',{\bf r})  n_{\mu}({\bf r}) d{{\bf r}'} d{\bf r},
     \label{Finteractionsingle}
\end{align}
where
\begin{align}
  u_{i}({\bf r}',{\bf r}) = - \left( {\bs}_{{\bf r}'} \cdot {\bf \nabla}_{{\bf r}'} \right) \left( {\bs}_{{\bf r}} \cdot {\bf \nabla}_{{\bf r}}  \right)  G_{1}({\bf r}',{\bf r})  \label{Potentialsingle}
\end{align}
depends on the single particle Green's function $G_{1}({\bf r}',{\bf r})$.

Rotating the effective colloidal particle (the solitonic texture enclosed within $\zeta$) amounts to a rigid body rotation of the distribution $n_{\mu}(a,\theta)$ about the $z-$axis passing trough the disk center. 
For the symmetric skyrmion, when the far field points out of the plane, there is no preferential direction for perturbations in the $(x,y)$ plane, and rotating $n_{\mu}(a,\theta)$ does change the free energy. 
However, there is an energetic cost associated with rotating the particle when the far field is not fully perpendicular to the $(x,y)$ plane, as is the case of the bimerons.
To get insight into the emerging elastic torque, we calculate the functional differential of Eq.~(\ref{Finteractionsingle}), which gives the difference between the free energy of a rotated and non-rotated particles (see Supplementary Note 2).
The resulting variation in the free energy can be expressed as
\begin{align}
  \delta U_{i} = & -4 \pi k \left(
  \frac{\Delta {\bf p}^{(\mu)} \cdot {\bf p}_{0}^{(\mu)}}{a^2}
  + 4 \frac{\Delta {\bf Q}^{(\mu)} : {\bf Q}_{0}^{(\mu)}}{a^4}
  \right),
  \label{Particlefarfieldint}
\end{align}
where $\Delta {\bf p}^{(\mu)}$ and $\Delta {\bf Q}^{(\mu)}$ are the differences between the rotated and non-rotated particle's dipole and quadrupole moments, while ${\bf p}_{0}^{(\mu)}$ and ${\bf Q}_{0}^{(\mu)}$ are the non-rotated equilibrium moments.
As mentioned above, $\delta U_{i}$ must be zero for the case of axially symmetric skyrmions, which means that $\Delta {\bf p}^{(\mu)}= 0$ and $\Delta {\bf Q}^{(\mu)}=0$  for all rotated states. However, for the bimeron ${\bf p}_{0}^{(\mu)}$ corresponds to the equilibrium dipole that can be different from the rotated one. ${\bf p}_{0}^{(\mu)}$ also characterises the director distribution far from the particle that minimizes the free energy and depends on the far field director \cite{brochard1970theory}. Therefore, equation (\ref{Particlefarfieldint}) shows the tendency of a rotated (transient) dipole to align with ${\bf p}_{0}^{(\mu)}$, or equivalently with the far field director. \newline

\noindent {\bf Elastic multipoles interaction} \newline

\noindent Effective interaction potential between two solitons $1$ and $2$, which are enclosed by two circular contours $\zeta_{1}$ and $\zeta_{2}$ centered at ${\bf O}_{1}$ and ${\bf O}_{2}$, respectively, can be obtained by replacing $n_{\mu}$ in Eq.~(\ref{FrankOseen_2D_Sk_one_cst}) with its expression in Eq.~(\ref{nmu}) as before, and considering that $n_{\mu}({\bf r}')$ and $n_{\mu}({\bf r})$ are the director distribution at contours $\zeta_{1}$ and $\zeta_{2}$.
Then, the two-particle interaction potential reads
\begin{align}
    U_{12} & =
     \frac{k}{2} \int_{\zeta_{1}} \int_{\zeta_{2}}
     n_{\mu}({\bf r}')
     u_{12}({\bf r}',{\bf r})  n_{\mu}({\bf r}) d{\bf r}' d{\bf r},
     \label{Finteraction}
\end{align}
where
\begin{align}
  u_{12}({\bf r}',{\bf r}) = - \left[ \left( {\bs}_{2} \cdot {\bf \nabla}_{{\bf r}'}  \right)  \left( {\bs}_{1} \cdot {\bf \nabla}_{\bf r} \right) + \left( {\bs}_{1} \cdot {\bf \nabla}_{{\bf r}'}  \right)  \left( {\bs}_{2} \cdot {\bf \nabla}_{\bf r} \right) \right] G_{2}({\bf r}',{\bf r})
  \label{Potential}
\end{align}
is expressed in terms of the two particle Green's function $G_{2}({\bf r}',{\bf r})$, and ${\bf \bs}_{1}$ and ${\bf \bs}_{2}$ denote the inward unit normal vectors to  $\zeta_{1}$ and $\zeta_{2}$, respectively. $G_{2}({\bf r}',{\bf r})=0$ for ${\bf r}' \in \zeta_1$ and ${\bf r}' \in \zeta_2$ and can only be calculated approximately (see Supplementary Note 3 for details).

We adopt the method developed by Pergamenshchik for colloids in three dimensions \cite{PhysRevE.76.011707,Pergamenshchik2007-gi}, which is based on a method of images. 
The result can be expressed in the form of an expansion in a power series of the small parameter $a/R$, where $R = \vert{\bf O}_{1}-{\bf O}_{2}\vert$ is the center-to-center distance, as follows 
% %

%
\begin{align}
    U_{12} = U_{12}^{q-p} + U_{12}^{p-p} + U_{12}^{q-Q} + U_{12}^{p-Q} + U_{12}^{Q-Q} + h.o.t.,
    \label{U12}
\end{align}
\begin{align}
    \frac{U_{12}^{q-p}}{2 \pi k} = - \frac{1}{R} \left[ 
    q_{1}^{(\mu)} \left( {\bf p}^{(\mu)}_{2} \cdot {\bf u} \right) - q_{2}^{(\mu)} \left( {\bf p}^{(\mu)}_{1} \cdot {\bf u} \right)
    \right],
    \label{Uqp}
\end{align}
\begin{align}
    \frac{U_{12}^{p-p}}{4 \pi k} = - \frac{1}{R^{2}} \left[
    {\bf p}_{1}^{(\mu)} \cdot {\bf p}_{2}^{(\mu)} - 2 \left(  {\bf p}_{1}^{(\mu)} \cdot {\bf u} \right) \left( {\bf p}_{2}^{(\mu)} \cdot {\bf u} \right) \right],
    \label{Upp}
\end{align}
\begin{align}
    \frac{U_{12}^{q-Q}}{4 \pi k} = - \frac{1}{R^{2}} \left\{
    q_{1}^{(\mu)} \left[ \left( {\bf Q}_{2}^{(\mu)} \cdot {\bf u} \right) \cdot {\bf u} \right] +
    q_{2}^{(\mu)} \left[ \left( {\bf Q}_{1}^{(\mu)} \cdot {\bf u} \right) \cdot {\bf u} \right]
    \right\},
    \label{UcQ}
\end{align}
\begin{align}
    \frac{U_{12}^{p-Q}}{12 \pi k} & = \frac{1}{R^{3}} \left\{\left[ \left( {\bf Q}_{1}^{(\mu)} \cdot {\bf p}_{2}^{(\mu)} \right) \cdot {\bf u} \right] - \left[ \left( {\bf Q}_{2}^{(\mu)} \cdot {\bf p}_{1}^{(\mu)} \right) \cdot {\bf u} \right],
    \right.
    \nonumber \\
    &
    \left. 
    \hspace*{1cm}
    +
    2\left( {\bf p}_{1}^{(\mu)} \cdot {\bf u} \right) \left[ \left( {\bf Q}_{2}^{(\mu)} \cdot {\bf u} \right) \cdot \bf{u} \right]  -
    2\left( {\bf p}_{2}^{(\mu)} \cdot {\bf u} \right) \left[ \left( {\bf Q}_{1}^{(\mu)} \cdot {\bf u} \right) \cdot \bf{u} \right] \right\},
    \label{UpQ}
\end{align}
\begin{align}
    \frac{U_{12}^{Q-Q}}{8 \pi k} & = - \frac{1}{R^{4}} \left\{
    {\bf Q}_{1}^{(\mu)} : {\bf Q}_{2}^{(\mu)} - 8 \left( {\bf Q}_{1}^{(\mu)} \cdot {\bf u} \right) \cdot \left( {\bf Q}_{2}^{(\mu)} \cdot {\bf u} \right)
    \right.
    \nonumber \\
    &
    \left.
    \hspace*{4.5cm}
    + 12 \left[ \left( {\bf Q}_{1}^{(\mu)} \cdot {\bf u} \right) \cdot {\bf u} \right] \left[ \left( {\bf Q}_{2}^{(\mu)} \cdot {\bf u} \right) \cdot {\bf u} \right],
    \right\}.
    \label{UQQ}
\end{align}
with ${\bf u} = {\bf R}/R$.
The above equations are the general results for any two interacting particles, expressed in terms of the individual particles' elastic multipole moments $q_{i}^{(\mu)}$, ${\bf p}_{i}^{(\mu)}$ and ${\bf Q}_{i}^{(\mu)}$, $i=1,2$ and $\mu = x,y$, as defined in Eqs.~(\ref{dipolemoment}) and (\ref{quadrupole}). At leading order, $U_{12}$ is obtained when elastic multipole moments are calculated by using one-particle director distributions, i.e. by neglecting effects of particle 2 on the director distribution on $\zeta_1$ and vice versa. \newline

\noindent {\bf Elastic dipole and quadrupole moments for skyrmions} \newline

\noindent In this section we compute the multipole moments for the skyrmions and bimerons numerically using director distributions which minimize the Frank-Oseen free energy (\ref{eq:free_energy}). As was mentioned above, the skyrmions are stable when the electric field is $off$, while the bimerons$-$when the field is $on$. To reflect this fact we denote the dipole moments for skyrmions and bimerons as ${\bf p}^{(\mu)}_{off}$ and ${\bf p}^{(\mu)}_{on}$, and their quadrupole moments as ${\bf Q}^{(\mu)}_{off}$ and ${\bf Q}^{(\mu)}_{on}$, respectively.

\begin{figure}[ht]
\centering
\includegraphics[width=1\columnwidth]{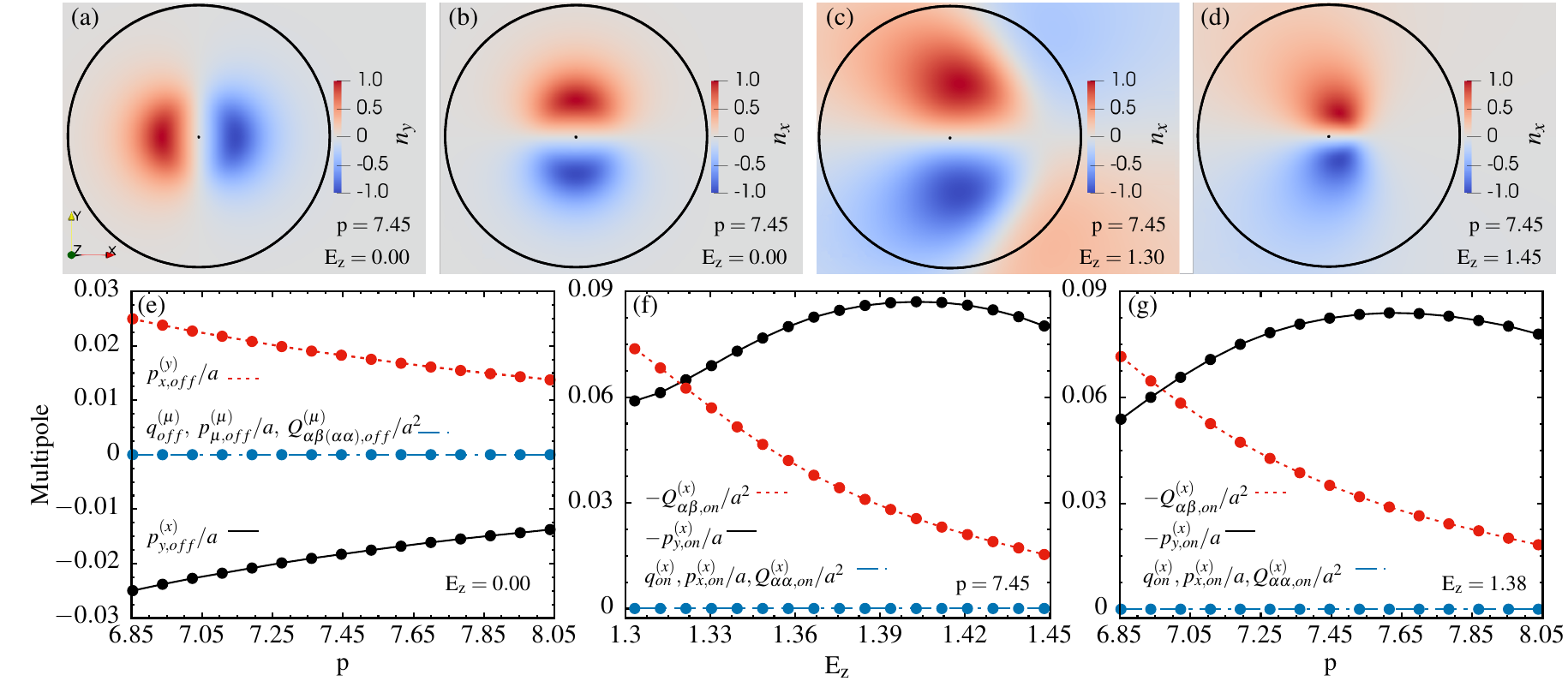} 
  \caption{{\bf Multipole moments for skyrmions and bimerons.} Colour coded $n_x$ (a) and $n_{y}$ (b) components of the director field corresponding to the skyrmion. (c) and (d) show colour coded $n_x$ for the bimeron at two values of the electric field as indicate in the panels. The cholesteric pitch ${\mathrm p}=7.45$ in all the cases. The black solid circles  represent the contour $\zeta$, which separates the inner soliton regions with $n_{\mu}\sim 1$ from the outer regions where $n_{\mu}\ll 1$. (e) Multipole moments calculated for the skyrmions as a function of the pitch. Due to symmetries of the director field, only dipole moments are different from zero. (f) and (g) shown the multipole moments for the bimeron as a function of the electric field and the pitch, respectively. For small $E_z$ and ${\mathrm p}$, the quadrupole moment dominates over the dipole, while the opposite is true for large $E_z$ and ${\mathrm p}$. $E_z$ is given in units of $\sqrt{W_0/\varepsilon_0 \Delta\varepsilon}$, and ${\mathrm p}$ in units of $\sqrt{k/W_0}$ where $k$ is the average elastic constant.
   }
  \label{FigMultipole}
\end{figure}
Figures~\ref{FigMultipole}(a) and (b) show the director components $n_x$ and $n_y$ for the skyrmion, and Figures~\ref{FigMultipole}(c) and (d) for the bimerons at two different values of the electric field strength $E_z$. The solitonic textures with large director deformations are enclosed within the solid circles which represents the contours $\zeta$ used to calculate the elastic multipole moments according to Eqs.~(\ref{dipolemoment}) and (\ref{quadrupole}). 
Figure~\ref{FigMultipole}(e) shows the resulting moments for the skyrmions, where due to the perturbation's symmetry (see Figs.~\ref{FigMultipole}(a) and (b)) only the elastic dipoles ${\bf p}^{(\mu)}_{off}$ are different from zero, and from the symmetry also follows that $p_{x,off}^{(y)} = - p_{y,off}^{(x)}$. The absolute value of the dipoles decreases with the increasing of the pitch,  which is related to the decreasing size of the skyrmion texture enclosed by $\zeta$. 

Elastic multipole moments associated with the bimerom texture are shown in 
Figs.~\ref{FigMultipole}(f) and (g) as functions of the electric field strength and the pitch.
In this case, only the elastic dipole in the $+\hat{{\bf y}}$ direction and the components $Q_{\alpha,\beta}^{(x)},~\alpha \neq \beta$, are different from zero.
The bimeron's quadrupole moment is small for strong electric fields as compared to the dipole moment due to the enhanced dipolar symmetry of the director perturbation as shown in Fig.~\ref{FigMultipole}(d). With the decreasing electric field 
the bimeron texture's size increases,
Fig.~\ref{FigMultipole}(c), and the quadrupole moment increases.
Similar effect is observed when decreasing the cholesteric pitch at fixed electric field, 
Fig.~\ref{FigMultipole}(g).
Summarising, the non-zero components of the elastic dipoles and quadrupoles for stable skyrmions (field $off$) and bimerons (field $on$) are

\begin{align}
    & p^{(y)}_{x,off} = -p^{(x)}_{y,off} \equiv p_{off},
    \nonumber \\
    & p^{(x)}_{y,on} \equiv - p_{on},
    \nonumber \\
    & Q^{(x)}_{\alpha \beta,on} = Q^{(x)}_{\beta \alpha,on} \equiv Q.
    \label{Multipole}
\end{align}
The first equation follows from the rotational symmetry of the skyrmion,
which also imposes vanishing quadrupole moments.

Using the above results for the multipoles in Eqs.~(\ref{Uqp})-(\ref{UQQ}), we find that $U_{12}^{q-p} = U_{12}^{q-Q} = 0$ because $q^{(\mu)}=0$ in all the cases. Additionally, by neglecting the rotation of the inner textures of solitons with respect to each other or the far field we find that $U_{12}^{p-Q}$ is also zero. Therefore, the only relevant effective interaction potentials are dipole-dipole ($U_{12}^{p-p}$) and quadrupole-quadrupole ones ($ U_{12}^{Q-Q}$). In principle, there also exist higher order elastic multipoles, but these are not considered in the present study.
\begin{figure}[ht]
  \centering
  \includegraphics[width=0.8\columnwidth]{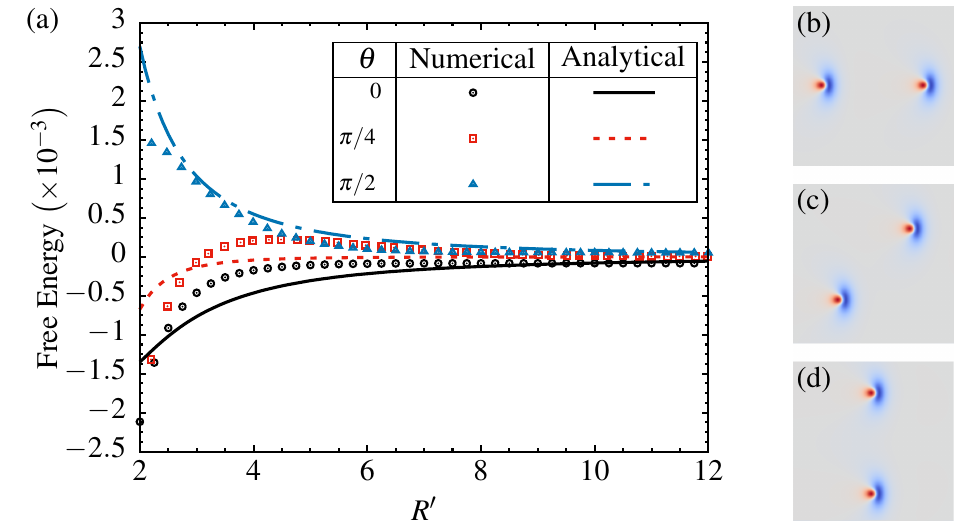} 
     \caption{{\bf Bimeron pairwise effective interaction potentials.} Effective interaction potential as a function of the separation $R’ = R/a$ between two bimerons calculated using Eq.~(\ref{U12}) with $p_{on}/a=0.09$ and $Q/a^{2}=0.03$,  solid lines, and via numerical minimization of the Frank-Oseen free energy  (\ref{eq:free_energy}).  The analytical potentials are given in units of $4\pi k$ and the numerical data in units of $2\pi k$, where $k$ is the average elastic constant. Different lines and symbols correspond to different angles $\theta$ between the center-to-center vector ${\bf R}$ and the $x$ axis. Numerical data is computed at $p=7.61$ and $E_{z}=1.41$, and  the analytical curves are  obtained as a sum  of the dipole-dipole and quadrupole-quadrupole potentials using $p_{on}/a=0.09$ and $Q/a^{2}=0.03$. $E_z$ is given in units of $\sqrt{W_0/\varepsilon_0 \Delta\varepsilon}$, and ${\mathrm p}$ in units of $\sqrt{k/W_0}$. (b)-(d) show colour coded $n_z$ component of the numerically obtained director fields calculated at $R=4a$ and $\theta =0,\pi/4, \pi/2$, other values of the parameter are the same as in (a). }
  \label{EnergyxR}
\end{figure}

Figure \ref{EnergyxR} compares for two bimerons, the analytical interaction potential (\ref{U12}), $U_{12}/4 \pi k$ (lines), with the numerically calculated Frank-Ossen free energy (\ref{eq:free_energy}), $F/2 \pi k$ (symbols) as functions of the distance between the  particles, and for several values of angle $\theta$ between the center-to-center vector ${\bf R}$ and the $x$ axis. The analytical profiles which are calculated using $p_{on}/a=0.09$ and $Q/a^{2}=0.03$  show a semi-quantitative agreement with the numerical results for large separations $R>6a$. The effective interaction is purely attractive when the bimerons are aligned “head-to-tail”,  $\theta=0$, while the bimerons repel each other at all distances for the “side-by-side” alignment, $\theta = \pi/2$. For oblique angles, the particles experience repulsive interaction at large and attractive one at short separations. The numerical data exhibit a pronounced free energy barrier at $R \approx 4$, which separates the attractive and repulsive regimes. The barrier is not visible on the analytical curve at this scale. \newline

\noindent {\bf Particle-based model for soliton self-assembly} \newline

\noindent The LC skyrmions subject to oscillating electric fields acquire transnational motion and exhibit reconfigurable self-assembling behaviour driven by out-of-equilibrium elastic forces \cite{Sohn2019}. In our recent work \cite{teixeira2024} we have developed a particle-based model of the translational motion of a skyrmion. The model is based on coarse-grained external time-dependent forces which describe the tendency of skyrmions to move forward or backward when the electric field is turned {\it on} and {\it off}. The functional form of these forces was motivated by fine-grained modeling in terms of the Frank-Oseen free energy. The coarse-grained model correctly described the dependence of the skyrmion velocity on the frequency and the duty cycle of the electric field. Here, we augment the particle-based model of \cite{teixeira2024} by including soliton interactions at the pairwise level as given by Eq.~(\ref{U12}).

According to the particle-based model, the soliton's velocity ${\bf v}$ satisfies, ignoring thermal fluctuations, the following dimensionless system of equations
\begin{align}
&
\dot{\bf v}(t) = 
- {\bf v}(t)
+ F_{n} ( t ) \hat{\bf v}(t) + {\bf F}_{int} (t),
\label{eq_of_motion} \\
&
F_{on}(t) = e^{-\alpha_{on} t},\quad F_{off}(t) = A_r e^{-\alpha_{off} t},
\label{active_force}
\end{align}
where the first term in the r.h.s of (\ref{eq_of_motion}) is an effective drag force, the second term is the active force that describes the forward/backward motion of solitons in the direction  $\hat{\bf v}(t)$ due to the non-reciprocal morphing of the LC director field in response to the changing electric field, and the last term encodes soliton interactions and is calculated as ${\bf F}_{int}=-{\bf \nabla}_{\bf R}U_{12}/k$. The dimensionless phenomenological parameters $\alpha_{on},\alpha_{off}$ and $A_r$ in the active force (\ref{active_force}) can be related to the strength of the electric field and the cholesteric pitch as discussed in \cite{teixeira2024}. We have implemented this particle-based model in the open source package "Large-scale Atomic/Molecular Massively Parallel Simulator" (LAMMPS) \cite{PLIMPTON1995,LAMMPS}, following our earlier implementations for active Brownian particles \cite{PhysRevE.106.024609,Dias2023}.

When the electric field is initially in $off$ state and then is suddenly turned $on$, the symmertic skyrmion texture morphs into the asymmetric bimeron one, and the associated dipole and quadrupole moments evolve from $p_{off} \rightarrow p_{on}$ and $0 \rightarrow Q$. The direction of the arrows in the above expressions reverses when the electric field changes from $on$ to $off$ state. For time-periodic electric fields the multipole moments evolves in time, and for long enough periods the dipole and quadrupole moments oscillates between the two limiting equilibrium values, i.e. $p \in [p_{off},p_{on}]$ and $Q \in [0,Q_{on}]$. We assume that the dipoles and quadrupoles depend exponentially on time similarly to the active force in (\ref{active_force}), which is a good approximation at low frequencies of the driving electric field. We also consider pulse-width modulated profile of the electric field when the field is $on$ for duration $t_{on}$ and $off$ for duration $t_{off}$. Then, within the $j-$th cycle (or period), when the field is changing from $off$ to $on$ the elastic dipoles and quadrupoles have the following dynamics

\begin{align}
    & p^{(y)}_{x}(t) = p_{off}~e^{-\alpha_{on} \left[t - (j-1) (t_{on}+t_{off})
    \right]},
    \nonumber \\
    & p^{(x)}_{y}(t) = - p_{off} + (p_{off} - p_{on}) \left(1 - e^{-\alpha_{on} \left[t - (j-1) (t_{on}+t_{off})
    \right]}\right),
    \nonumber \\
    & Q^{(x)}_{\alpha \beta} (t) = Q_{on} \left( 1 - e^{-\alpha_{on} \left[t - (j-1) (t_{on}+t_{off}),
    \right]} \right),
    \label{multipoleon}
\end{align}
and when the field is changing from $on$ to $off$ we have 
\begin{align}
    & p^{(y)}_{x}(t) = p_{off} \left( 1 - e^{-\alpha_{off} \left[t - j~t_{on} - (j-1) t_{off}
    \right]} \right),
    \nonumber \\
    & p^{(x)}_{y}(t) = - p_{on} + (p_{on} - p_{off}) \left( 1 - e^{-\alpha_{off} \left[t - j~t_{on} - (j-1) t_{off}
    \right]} \right),
    \nonumber \\
    & Q^{(x)}_{\alpha \beta} (t) = Q~e^{-\alpha_{off} \left[t - j~t_{on} - (j-1) t_{off}
    \right]}.
    \label{multipoleoff}
\end{align}
\begin{figure}%[Ht]
\centering
\includegraphics[width=1.0\columnwidth]{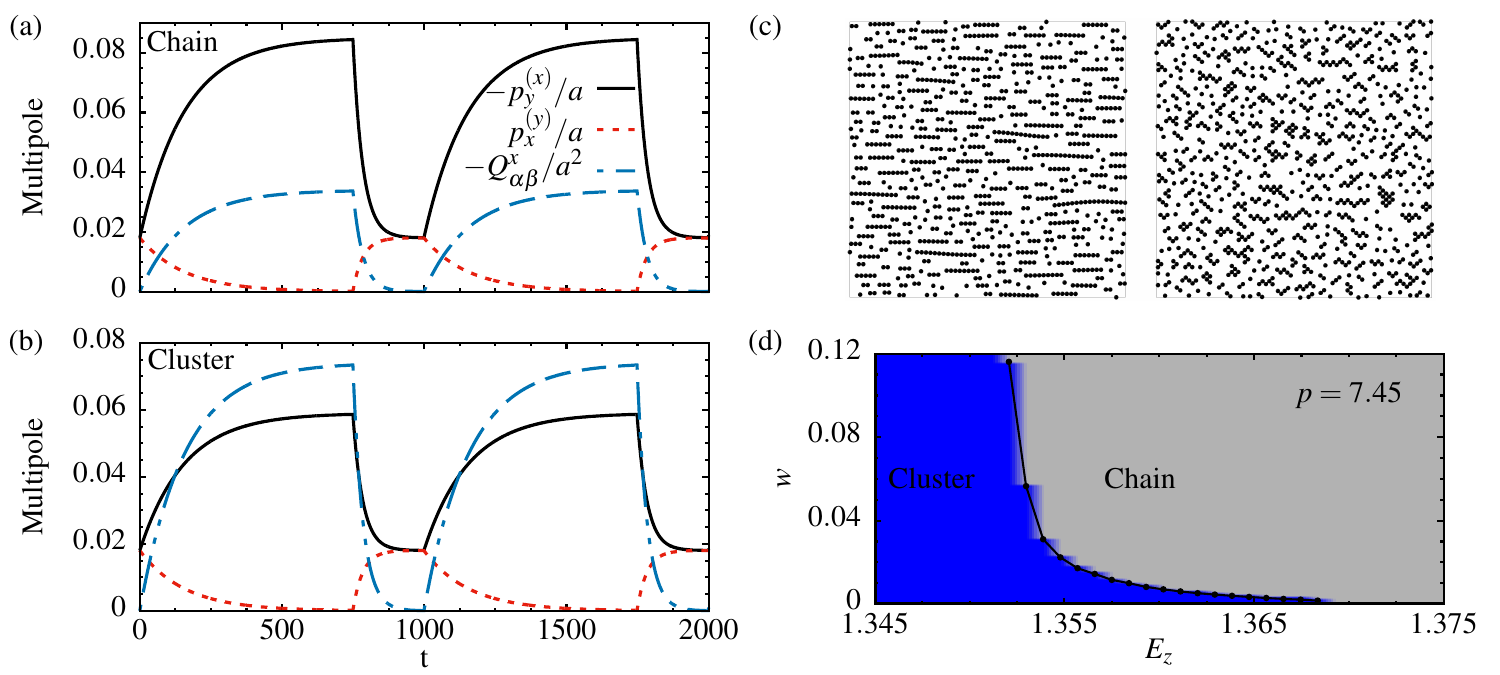} 
  \caption{{\bf Particle-based model simulations of soliton self-assembly.} (a) and (b)  dipole and quadrupole moments calculated according to (\ref{multipoleon}) and (\ref{multipoleoff}) as functions of time given in units of $\gamma/m$, where $m$ is an effective skyrmion mass and $\gamma$ is a friction parameter,  see ref.~\cite{teixeira2024} for more details. In (a) $p_{on}/a = 0.085$, $p_{off}/a = 0.018$, and $Q/a^2 = - 0.034$, which corresponds to $E_z = 1.375$; and in (b) $p_{on}/a = 0.059$, $p_{off}/a = 0.018$, and $Q/a^2 = - 0.074$, which corresponds to $E_z = 1.30$, in (a) and (b) the pitch ${\mathrm p} = 7.45$. $E_z$ is given in units of $\sqrt{W_0/\varepsilon_0 \Delta\varepsilon}$, and ${\mathrm p}$ in units of $\sqrt{k/W_0}$. The frequency of the pulse-width modulated electric field $w=10^{-3} m/\gamma$; the duty cycle of the field $t_{on}/t_{off} = 0.75$. (c) Typical molecular dynamics simulations snapshots showing the formation of soliton chains (left) and rhomboidal clusters (right). The snapshots contain 1000 particles in a square domain with the lateral size $125a$ and are taken at $t=10000\gamma/m$; $A_{r}=3.02$, $\alpha_{on}=6.4 \times 10^{-3}$, and $\alpha_{off}/\alpha_{on}=4.50$. The periodic boundary conditions are applied in both directions and the simulations are initiated with random configurations of solitons. The left/right panel of (c) corresponds to the multipole dynamics of (a)/(b). (d) A ground state phase diagram in the $(E_z,w)$ plane showing the stability regions of soliton chains and clusters. The potential energy per particle is compared for chains and 2D clusters with 113 particles in both cases, by using pair potential \eqref{Uij} together with  Eq.~(\ref{eq:multipoles_on_long-time}) for the dipole and quadrupole moments. The values of $p_{on}$, $p_{off}$ and $Q$ in (\ref{eq:multipoles_on_long-time}) for different  $E_z$ and at ${\mathrm p}=7.45$ are calculated numerically, the corresponding results are plotted in Fig.~\ref{FigMultipole}(f).}
  \label{Multipolextime}
\end{figure}
Figure~\ref{Multipolextime}(a) shows the time evolution of the elastic multipoles calculated according to equations (\ref{multipoleon}) and (\ref{multipoleoff}). The dipole moment at the end of the field $on$ state $p_{on}/a = 0.06$ and the dipole-dipole interaction dominates the quadrupole-quadrupole one with $Q/a^2 = - 0.02$. This case corresponds to strong electric fields when the dipole is more relevant than the quadrupole leading to the formation of soliton chains aligned in the $x$ direction as is illustrated in the  left panel of Fig.~\ref{Multipolextime}(c). Figure ~\ref{Multipolextime}(b) corresponds to weaker electric fields, when at the end of the field $on$ state the dipole moment $p_{on}/a = 0.05$ is less relevant than the quadrupole with $Q/a^2 = - 0.08$. The dominating quadrupole-quadrupole interaction promotes soliton self-assembly into crystal-like clusters with the rhomboidal unit cell, see the right panel in Fig.~\ref{Multipolextime}(c). In both cases the dipole moment at the end of the field $off$ state $p_{off}/a = 0.015$, which according to the analysis of ref.~\cite{teixeira2024} corresponds to maintaining the cholesteric pitch fixed. The field-driven transition between soliton chains and cluster has also been observed experimentally \cite{Sohn2019}. In the next subsection we calculate a ground state chain-cluster phase diagram in the plane $(E_z,w)$, where $w$ is the frequency of the electric field. \newline

\noindent {\bf Chain-cluster ground state phase diagram} \newline

\noindent Two particles, say $1$ and $2$, which are non-rotated (in a sense of section {\it Elastic torque on an asymmetric skyrmion}) with respect to the far field director, interact only via the dipolar, Eq.~(\ref{Upp}), and the quadrupolar, Eq.~(\ref{UQQ}) coupling. In time-dependent electric fields the dipole and quadrupole moments evolve according to Eqs.~(\ref{multipoleon}) and (\ref{multipoleoff}), resulting in three contributions to the total interaction potential: the dipole-dipole interactions $U_{12}^{p-p}$ where the particles $1$ and $2$ have identical dipole moments $p_1=p_2=p_{y}^{(x)}(t)$ or $p_1=p_2=p_{x}^{(y)}(t)$, and the quadrupole-quadrupole interaction $U_{12}^{Q-Q}$ with the identical quadrupole moments $Q_{\alpha \beta}^{(x)}(t)$.
The sum of the three contributions gives
\begin{align}
  U_{12}(R,\theta) =  
  &\frac{4 \pi k}{R^2} \left[
  \frac{12 \left [ Q_{\alpha \beta}^{(x)}(t) \right ]^{2}}{R^2} \cos{\left(4 \theta \right)}
  - \left( \left [p_{y}^{(x)}(t)\right ]^{2} - \left [ p_{x}^{(y)}(t)\right ] ^{2} \right) \cos{\left(2 \theta \right)}
  \right].
  \label{Uij}
\end{align}
When the dipolar interactions dominate, $U_{12}$ in (\ref{Uij}) attains global minimum (at fixed $R$) at $\theta = 0,\pi$, which facilitates the formation of chains oriented perpendicular to the far field director. On the other hand, when the quadrupole moment dominates, the global minima (at fixed $R$) occur at $\theta=n\pi/4$, with $n$ an integer number, favouring cluster formation with diagonal bounds, as shown in the right snapshot of Fig.~\ref{Multipolextime}(c).

We estimate a ground state chain-cluster phase diagram in the electric field, 
 frequency $(E_z,w)$ plane by comparing the potential energy per particle of closely packed  soliton aggregates. We consider chains oriented perpendicular to the far field director, and 2D clusters with the particles bonds oriented at $45^{\circ}, 135^{\circ}$ with respect to the $x$ axis. The multipole dynamics in equations~(\ref{multipoleon}) and (\ref{multipoleoff}) assume low frequencies, or long $t_{on}$ and $t_{off}$, when the moments relax completely to the values in (\ref{Multipole}) at the end of each $on$ and $off$ cycles. At higher frequencies, we account for a partial relaxation of the moments by modifying appropriately the r.h.s. of Eqs.~(\ref{multipoleon}) and (\ref{multipoleoff}), similarly to Eq.~(19) of ref.~\cite{teixeira2024}. After an initial, frequency dependent transient the moments attain a steady periodic dynamics as shown in figure 9(a) of ref.~\cite{teixeira2024}. Finally, because the cluster formation is favoured by the quadrupolar interaction, which is the strongest at the end of each $on$ cycle, we use the asymptotic $t\rightarrow \infty$ values of the dipole and quadrupole moments at the end of the $on$ cycle in the calculation of the agregates potential energies. 
In the limit of $t \rightarrow \infty$, the corresponding multipole moments at the end of the $on$ cycle read
\begin{align}
    & p^{(y)}_{x}(t) = p_{off}~{\cal C},
    \nonumber \\
    & p^{(x)}_{y}(t) = - p_{off} + (p_{off} - p_{on}) \left(1 - ~{\cal C}\right),
    \nonumber \\
    & Q^{(x)}_{\alpha \beta} (t) = Q \left( 1 - {\cal C} \right),
    \label{eq:multipoles_on_long-time}
\end{align}
where
\begin{align}
  {\cal C} \equiv 
  \frac{ e^{\alpha_{off} t_{off}} - 1}{e^{\alpha_{on} t_{on} + \alpha_{off} t_{off}}-1}.
\end{align}
Using the above equations in (\ref{Uij}) and comparing the potential energies per particle for clusters and chains, the resulting phase diagram is shown in Fig.~\ref{Multipolextime}(d). For weak electric fields the quadrupole moment is larger than the dipole one (Fig.~\ref{FigMultipole}(f)) and the formation of clusters is observed.
For stronger electric fields, chains are favored by the stronger dipolar interaction. For intermediate values of the electric field, however, the dominance of dipolar and quadrupolar interaction depends on the frequency $w$: for high $w$ the dipolar interaction can be more relevant than the quadrupolar one,  while for low frequencies the opposite is true. \newline

\noindent {\Large \bf Conclusion} \newline

\noindent We have adapted a method of nematostatics, proposed by Pergamenshchik and Uzunova in ref. ~\cite{Pergamenshchik2007-gi} for nematic colloids, to derive analytical pair potentials for LC skyrmions. The method is based upon a multipole expansion, Eq.~(\ref{perturbation}), of the perturbations of the far field director due to the presence of 2D skyrmion textures. We have consider symmetric skyrmion and electric field deformed bimeron configurations. The director perturbations are quantified in terms of elastic multipole moments: elastic charge and dipole in Eq.~(\ref{dipolemoment}), and quadrupole in Eq.~(\ref{quadrupole}) moments. Thus, the symmetric skyrmion has only an elastic dipole, see Fig.~\ref{FigMultipole}(e), while the assymetric bimeron structure acquires dipole and quadrupole moments, see Figs.~\ref{FigMultipole}(f) and (g). In this study we have not considered higher, e.g. octupole, hexadecapole, etc. moments, because the main aim was on developing of the minimalist model for the out-of-equilibrium collective dynamics of skyrmions. The extension of the presented model, by including,  higher order multipoles will be published elsewhere.  

We have derived the leading contributions to the soliton-soliton effective potential $U_{12}$, which is presented as a sum (\ref{U12}) of terms describing the interactions between elastic multipoles, e.g. (\ref{Upp})/(\ref{UQQ}) is the interaction between two dipoles/quadrupoles. In molecular dynamics simulations, we have used the values of the multipole moments obtained from numerical director fields of isolated solitons, i.e. neglecting "polarisation" effects when the director field around a particle is perturbed by the presence of another particle. At this level of approximation, the identical skyrmions (LC director configuration at zero electric field) do not interact, which follows from adding up terms in (\ref{Upp}) for $\mu=x$ and $\mu=y$. Preliminary analysis shows that upon taking into account the effect of particle 2 on the director field around particle 1, the identical symmetric skyrmions do experience isotropic repulsion. The leading contribution stems from a coupling of the bare elastic dipole of particle 1 to an induced elastic charge on particle 2.  

The morphing of the director field upon turning the voltage on and off results in a temporal evolution of the elastic multipole moments. Their dynamics has been postulated in a simple exponential (relaxation) form \cite{teixeira2024}, suggested by the full numerical minimization of the Frank-Oseen free energy. The time-dependent moments determine pair interaction potentials between the particles of the coarse-grained model under dynamical conditions. Next, we have carried out molecular dynamics simulations of solitons' self-assembling behaviour showing reconfigurabale, with the frequency of the electric field, formation of chains and 2D clusters. The ground state configurational diagram in the $(E_z,w)$ plane highlights the frequency-driven chain-to-cluster reconfiguration. 

The developed approach comprises a minimal model of the LC skyrmions' collective behaviour in out-of-equilibrium setups. We have restricted our analysis to 2D systems, and considered only the lowest order elastic multipoles. We have also neglected the effects of the polarisation due to the presence of particle 2 when calculating the elastic multipoles of particle 1. Despite these approximations, the model captures essential experimental observations, such as chain-to-cluster transitions controlled by either the amplitude or the frequency of the external electric field. \newline  

\noindent {\Large \bf Methods} \newline

\noindent We minimize the Frank-Oseen elastic free energy \eqref{eq:free_energy} augmented by effective anchoring term in \eqref{fw} by using relaxation methods, which amounts to balancing elastic forces and dissipative forces due to the rotational dynamics of the director field. This renders the following dynamical equation for the director field $\nvec(t,\rvec)$ 
\begin{equation}
 \partial_t n_j(t,\rvec) = -\frac{1}{\gamma_1} \frac{\delta }{\delta n_j} \int_{L^2} \Big(u + u_{w} \Big)d^2r ,
 \label{director-time-eq}
\end{equation}
where $\gamma_1$ is the rotational viscosity, the expressions for the free energy densities $u$ and $u_w$ are defined in Eqs.~\eqref{eq:free_energy} and \eqref{fw}, respectively. The integral in the r.h.s. is taken of the two dimensional domain $L\times L$, taken to be within the $(x,y)$ plane, and gives the Frank-Oseen free energy per unit length along the $z-$direction. We apply periodic boundary conditions in the $x$ and $y$ directions. Equation (\ref{director-time-eq}) is solved subject to the constraint $(\nvec \cdot \nvec) = 1$. The spatial derivatives on the r.h.s. of Eq.~(\ref{director-time-eq}) are estimated via second order finite-differences and the integration over time is performed using the fourth-order Runge-Kutta method. The values of the model parameters used in this study are provided in table \ref{table}.

\begin{table}[h]
\caption{Parameters used in numerical integration of Eq.~(\ref{director-time-eq}) and physical units.}\label{tab1} 
\footnotesize
\begin{tabular}{@{}|l|l|l|l|}
\hline
symbol&sim. units & physical units&description\\
\hline
$\Delta x$&1 & 0.3125 $\mu$m& lattice spacing\\
$\Delta t$&1 & 10$^{-6}$ s& time step\\
$L$&300, 600 & $\approx 94, 188 \mu m$ & simulation box side length\\
$k_{1}$&17.2 &17.2$\times 10^{-12}$ N & splay elastic constant\\
$k_{2}$&7.51 &7.51$\times 10^{-12}$ N & twist elastic constant\\
$k_{3}$&17.2 &17.2$\times 10^{-12}$ N & bend elastic constant\\
$\gamma_1$&162&0.162 Pa s&director rotational viscosity\\
$\Delta\varepsilon$ & -3.7 & -3.7& dielectric anisotropy\\
\hline
\end{tabular}
\label{table}
\end{table}
\normalsize

As an initial conditions for numerical integration of Eq.~(\ref{director-time-eq}) we use the following {\it Ansatz}:
\begin{align}
 &n_x(\rvec) = \sin\left(a(\rvec) \right) \sin\left (sb(\rvec) + g \right)\nonumber\\
 &n_y(\rvec) = \sin\left(a(\rvec) \right) \cos\left (sb (\rvec) + g \right )\nonumber\\
 &n_z(\rvec) = -\cos\left(a(\rvec) \right),
 \label{toron_ansatz}
\end{align}
where 
\begin{equation}
 a (\rvec) =  \frac{\pi}{2}\left[1 - \tanh\left(\frac{B}{2}(\rho- \rho_0 \right)\right],
\end{equation}
\begin{eqnarray}
 && b(\rvec) = \tan^{-1}\left(\frac{x-C_x}{y-C_y}\right),\\
 && \rho=\sqrt{(x-C_x)^2+(y-C_y)^2} .
\end{eqnarray}
Here $(C_x,C_y)^{\mathrm{T}}$ defines the coordinates of the {\it Ansatz} centre, $\rho_0$ controls the size of the skyrmion, $B$ controls the width of the twisted wall of the skyrmion, $s$ is the winding number of the skyrmion, $g$ controls the way the director twists along the radial directions from the skyrmion center, $\rho$ is the distance from a given point $(x,y)^{\mathrm{T}}$ to the skyrmion center, and $b$ is the $2D$ polar angle. The values of the parameters used in the simulation are (in simulation units): $s=1$, $g = \pi/2$, $\rho_0=0.45{\mathrm p}$, $B=0.5$, $C_x=C_y=L/2$.

Before applying an external electric field $(0,0,E_z)^{\mathrm{T}}$, we brake "by hand" the axial symmetry of the {\it Ansatz} in Eq.~(\ref{toron_ansatz}). We start with $\Evec=(0,E_0,E_0)^{\mathrm{T}}$, and carry out 500 Runge-Kutta iterations on discretised Eq.~(\ref{director-time-eq}). This tilts the far field director along the $x-$axis. Next, we switch the field off and relax the director field for 5500 time steps, the resulting skyrmion configuration will remain slightly asymmetric, which is crucial to achieve the bimeron configuration.

\begin{figure}[ht]
  \centering
  \includegraphics[width=0.6\columnwidth]{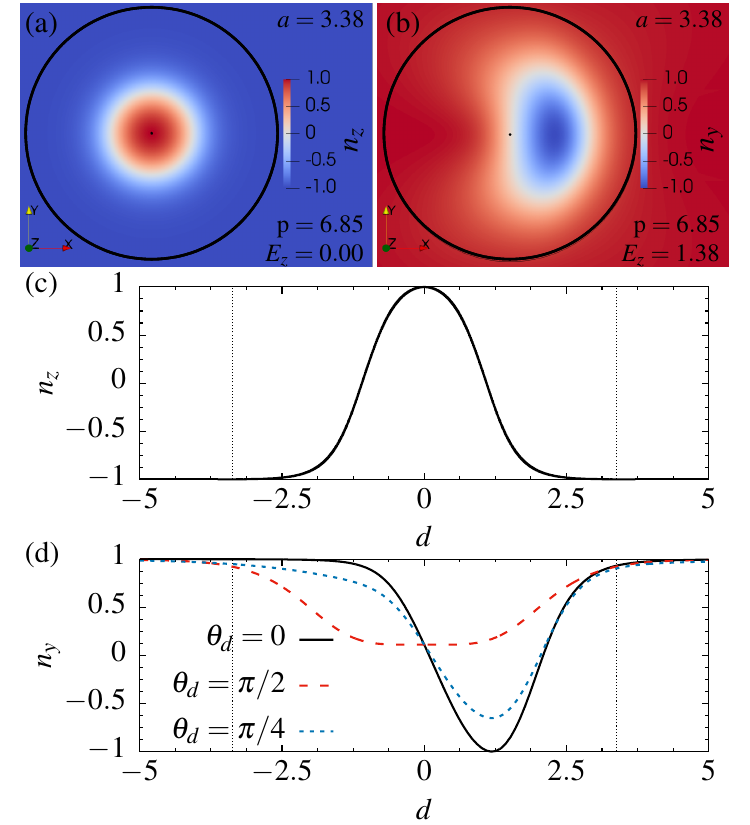} 
  \caption{{\bf Director perturbations for solitons.} (a) colour coded $z-$component $n_z$ of the skyrmion at zero electric field; (b) colour coded $y-$component $n_y$ of the bimeron at $E_z = 1.38$; the cholesteric pitch $\mathrm{p}=6.85$ in both cases; $E_z$ is given in units of $\sqrt{W_0/\varepsilon_0 \Delta\varepsilon}$, and ${\mathrm p}$ in units of $\sqrt{k/W_0}$. For the skyrmion, the far field director equals $\hat{z}$, while it equals $\hat{y}$ for the bimeron. The black circle of radius $a=3.38$ is centred at $(x_c,y_c)$ as defined in Eq.~(\ref{toron-center-of-mass}). (c) $n_z$ for the skyrmion as a function of the distance from the skyrmion's centre of mass $d=\sqrt{(x-x_c)^2+(y-y_c)^2}$. (d) $n_y$ for the bimeron as a function of $d$ for several values of the angle $\theta_d$ the vector $(x-x_c,y-y_c)^{\mathrm T}$ forms with the $x$ axes. The dotted vertical lines delimit the non-linear inner regions from the outer linear ones. }
  \label{Radius}
\end{figure}

The elastic multipole moments of a soliton with the dividing circle of radius $a$ which enclose  the region of large director deformation is calculated from the configuration $n_{\mu}(r=a,\theta)$ with the help of equations (\ref{dipolemoment}) and (\ref{quadrupole}). Figures \ref{Radius}(a) and \ref{Radius}(b) show the typical configurations of a skyrmion in the absence of electric field and a bimeron at $E_{z}=1.38\sqrt{W_0/\varepsilon_0 \Delta\varepsilon}$, respectively; the dividing circle is depicted by black solid circle and in both cases the cholesteric pitch is $\mathrm{p}=6.85\sqrt{k/W_0}$.  
The solid black circles of radius $a=3.38\sqrt{k/W_0}$ separate the non-linear inner regions from the linear outer ones, and are used to calculate the elastic multipole moments. The circle is centred in the "centre of mass" of the soliton, defined as
\begin{equation}
(x_c,y_c) = \frac{\int_{\Omega}n_{z}(\mathbf{r})(x,y) dA}{\int_{\Omega} n_{z}(\mathbf{r}) dA},  
\label{toron-center-of-mass}
\end{equation}
where the integration is over the domain defined by $n_{z}({\mathbf{r}}) \ge 0.3$.
 Figures~\ref{Radius}(c) and (d) shown the rationale why the value $a=3.38$ has been chosen for the radius of the dividing circle. Indeed, we find that for the skyrmion $|n_z(r) - 1|<0.0013$, Fig.\ref{Radius}(c), and for the bimeron $|n_y(r,\theta) - 1| < 0.1$, Fig.\ref{Radius}(d), when $r>3.38\sqrt{k/W_0}$. We note, that the actual size of the large deformation texture decreases when $\mathrm{p}$ and $E_{z}$ increase.

\vspace*{0.5cm}

\noindent {\Large \bf Acknowledgements} \newline

\noindent We acknowledge financial support from the Portuguese Foundation for Science and Technology (FCT) under Contracts no. PTDC/FIS-MAC/5689/2020 (https://doi.org/10.54499/PTDC/FIS-MAC/5689/2020), EXPL/FIS-MAC/0406/2021, CEECIND/00586/2017, UIDB/00618/2020 UIDB/00618/2020 (DOI 10.54499/UIDB/00618/2020),  and UIDP/00618/2020 UIDP/00618/2020 (DOI 10.54499/UIDP/00618/2020).\newline

\noindent {\Large \bf References}\newline

\bibliography{bibliography}

\end{document}